\documentclass[a4paper,fleqn,usenatbib]{mnras}

\usepackage[T1]{fontenc}
\usepackage{ae,aecompl}

\usepackage{graphicx}	
\usepackage{amsmath}	
\usepackage{amssymb}	
\usepackage{color}
\usepackage{aas_macros}
\usepackage{longtable}
\usepackage{tabulary}

\title[SXP4.78]{ The SMC X-ray binary SXP4.78 : a new Type II outburst and the identification and study of the optical counterpart}
\author[Monageng et al.]
  {I.M. Monageng$^{1}$\thanks{E-mail: itu@saao.ac.za}, 
  M.J. Coe$^{2}$,
  L.J. Townsend$^3$,
  D.A.H. Buckley$^1$,
  V.A. McBride$^1$,
  \newauthor 
  P.D. Roche $^4$, 
  J.A. Kennea$^5$, 
  A. Udalski$^6$, 
  P.A. Evans$^7$
\\
$^1$South African Astronomical Observatory, P.O Box 9, Observatory, 7935, Cape Town, South Africa\\
$^2$Physics \& Astronomy, University of Southampton, SO17 1BJ, UK\\
$^3$Department of Astronomy, University of Cape Town, Private Bag X3, Rondebosch 7701, South Africa\\
$^4$School of Physics \& Astronomy, Cardiff University, The Parade, Cardiff, CF24 3AA\\
$^5$Department of Astronomy and Astrophysics, The Pennsylvania State University, University Park, PA 16802, USA\\
$^6$Warsaw University Observatory, Al. Ujazdowskie 4, 00-478 Warszawa, Poland\\
$^7$University of Leicester, X-ray and Observational Astronomy Research Group, Leicester Institute for Space and Earth Observation, \\
Department of Physics \& Astronomy, University Road, Leicester LE1 7RH, UK\\
}

\date{Accepted XXX. Received YYY; in original form ZZZ}

\pubyear{2019}

\def\LaTeX{L\kern-.36em\raise.3ex\hbox{a}\kern-.15em
    T\kern-.1667em\lower.7ex\hbox{E}\kern-.125emX}

\def\g09{G10}

\def\lsim{\mathrel{\hbox{\rlap{\hbox{\lower4pt\hbox{$\sim$}}}\hbox{$<$}}}}
\def\gsim{\mathrel{\hbox{\rlap{\hbox{\lower4pt\hbox{$\sim$}}}\hbox{$>$}}}}

\makeatletter
\newcommand*{\rom}[1]{\expandafter\@slowromancap\romannumeral #1@}
\newcommand*{\Rom}[1]{\expandafter\@slowromancap\romannumeral #1@}
\makeatother

\begin{document}

\label{firstpage}
\pagerange{\pageref{firstpage}--\pageref{lastpage}}
\maketitle

\begin{abstract}
SXP4.78 was originally discovered in 2000 as a pulsar in the Small Magellanic Cloud (SMC) by the Rossi X-ray Timing Explorer (RXTE,) but it was not spatially located at that time. A new detection in 2018 with the Neil Gehrels \textit{Swift} Observatory during a Type \rom{2} outburst permitted its position to be accurately located and its optical counterpart identified. We report X-ray and optical monitoring covering epochs before and during the outburst. Using photometric data we show the long-term variability of the Be disc where we present flux and colour changes associated with the disc growth and decay over a period of $\sim$6000~days. We show evidence of disc growth during the recent outburst through an increase in the H$\alpha$ equivalent width and photometric flux. Period analysis was performed using both optical photometric and spectroscopic data, but with no significant detection of an orbital period. A modest periodic signature of 2.65~days was detected from the OGLE $I$ band data, however, but we attribute that to the non-radial pulsations (NRPs) of the Be star. We also obtained a blue spectrum from the Southern African Large Telescope (SALT) which permits us to classify the spectral type as B0.5 IV-V.
\end{abstract}


\section{Introduction}
\label{sec:introduction}
High mass X-ray binaries (HMXBs) are binary star systems comprising a compact object (neutron star or black hole) and a massive early-type companion (O or B spectral type). Based on the luminosity class of the massive companion HMXBs are generally divided into supergiant X-ray binaries (sgXBs; luminosity class \Rom{1} and \Rom{2}) and Be X-ray binaries (BeXBs; luminosity class \Rom{3}, \Rom{4} and \Rom{5}). BeXBs, which make up the largest subclass of the HMXB systems (49\% \citealt{2013ApJ...764..185C}), have a companion star with a geometrically thin Keplerian disc and a neutron star (NS; only one black hole system has been confirmed to date; \citealt{2014Natur.505..378C}). The disc variability is traced through the variability of the Balmer emission lines in the optical spectra, the strongest and best-studied of which is the H$\alpha$ line. The X-ray behaviour of BeXBs is characterised by two types of outburst events, namely type \Rom{1} ($L \leq 10^{37}$erg.s$^{-1}$) and type \Rom{2} ($L \geq 10^{37}$erg.s$^{-1}$; \citealt{1986ApJ...308..669S}). Type \Rom{1} outbursts occur regularly and are typically separated by the orbital period. The origins of type~\Rom{2} outbursts have remained elusive, with several models proposed for them (e.g. \citealt{2013PASJ...65...83M,2014ApJ...790L..34M,2017MNRAS.464..572M}). A comprehensive review of BeXBs is given in \citet{2011Ap&SS.332....1R}.

SXP~4.78, the subject of this paper, was reported as a new X-ray transient source in the Small Magellanic Cloud (SMC) from observations obtained with the Neil Gehrels \textit{Swift} Observatory (\citealt{2004ApJ...611.1005G}; as Swift J005139.2$-$721704) when it reached a luminosity of $4 \times 10^{37}$~erg.s$^{-1}$ \citep{2018ATel12209....1C}. Analysis of NICER and \textit{Fermi}/GBM data revealed that the source is associated with a known pulsar, XTE J0052$-$723 , with a pulse period of 4.78~s \citep{2018ATel12222....1S}. XTE J0052--723 was first discovered on 27 Dec 2000 (JD 2451906) during regular RXTE monitoring of the SMC (\citealt{2001IAUC.7562....1C}; \citealt{2003MNRAS.339..435L}). During this discovery outburst, there were 7 PCA detections of a strong 4.78\,s periodicity. A set of slew observations were used to localise the source of the pulsations to RA (J2000) = 00:52:17, Dec (J2000) = -72:19:51, which we now know to be several arc-minutes away from the actual source position (RA = 00:51:39.2, Dec = -72:17:03.6\citealt{2018ATel12209....1C}).

The spectrum is described by an absorbed blackbody and power-law model best-fitted by $N_H = 0.56 \pm 0.04 \times 10^{22}$~cm$^{-2}$, $kT = 0.120 \pm 0.03$~keV and $\Gamma = 1.07 \pm 0.03$ \citep{2018ATel12219....1G}. The refined position of the pulsar revealed an association with the B-star from Massey's catalogue [M2002] SMC~20671 \citep{2002ApJS..141...81M}, making the source a newly discovered BeXB. \citet{2018ATel12224....1M} performed spectral classification and conclude that the spectral type of the donor is B1-2e. The authors also report the presence of the H$\alpha$ and H$\beta$ lines in emission, which have previously been reported to be absent \citep{2004PASP..116..909E}. Period analysis of OGLE~\Rom{3} and \Rom{4} data showed a strong signal at 1.805~days \citep{2018ATel12229....1C}. It was also noted from the OGLE data that the $I$-band magnitude reached its brightest state in $\sim$6000~days, suggesting a recent growth in the circumstellar disc size associated with the X-ray activity. In this paper we present multiwavelength observations of SXP4.78, where we study its recent behaviour using data from SALT, \textit{Swift}, OGLE, RXTE and LCO.

\section{Observations}
\label{sec:observations}

\subsection{Swift}

\begin{figure*}
\includegraphics[width=18cm]{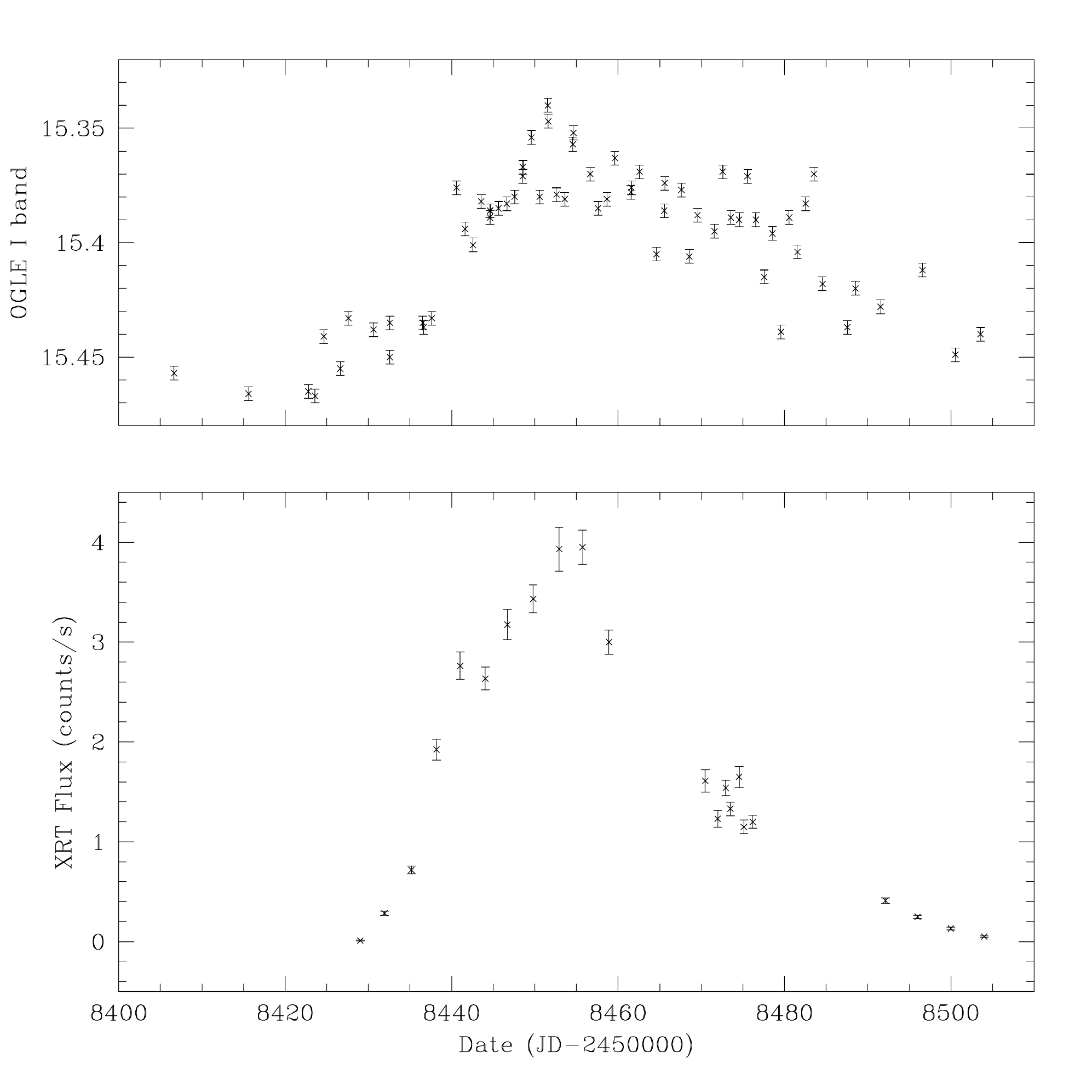}
\caption{The $I$-band and X-ray measurements of SXP 4.78 during the current outburst. The OGLE data are shown in the top panel and the Swift X-ray measurements are in the lower panel.}
\label{fig:xi}
\end{figure*}

The source was discovered by \textit{Neil Gehrels Swift Observatory} \citep{2004ApJ...611.1005G} whilst observing another SMC transient system, SXP 91.1 \citep{2018ATel12209....1C}. By good fortune SXP 4.78 lies only $\sim$5 arcminutes from SXP 91.1 and hence was comfortably in the same XRT field of view. The source was then monitored over the 0.3 -- 10 keV range throughout the outburst. The observations cover a couple of months and typical exposure times used were $\sim$ 0.9--1.1 ks, depending upon the spacecraft visibility window at that time. The XRT lightcurve was produced following the instructions described in the Swift data
analysis guide (http://www.swift.ac.uk/analysis/xrt/). The results are shown in Fig.~\ref{fig:xi}.

\subsection{SALT}
\begin{figure*}
\resizebox{\hsize}{!}
            {\includegraphics[angle=-90,width=17cm]{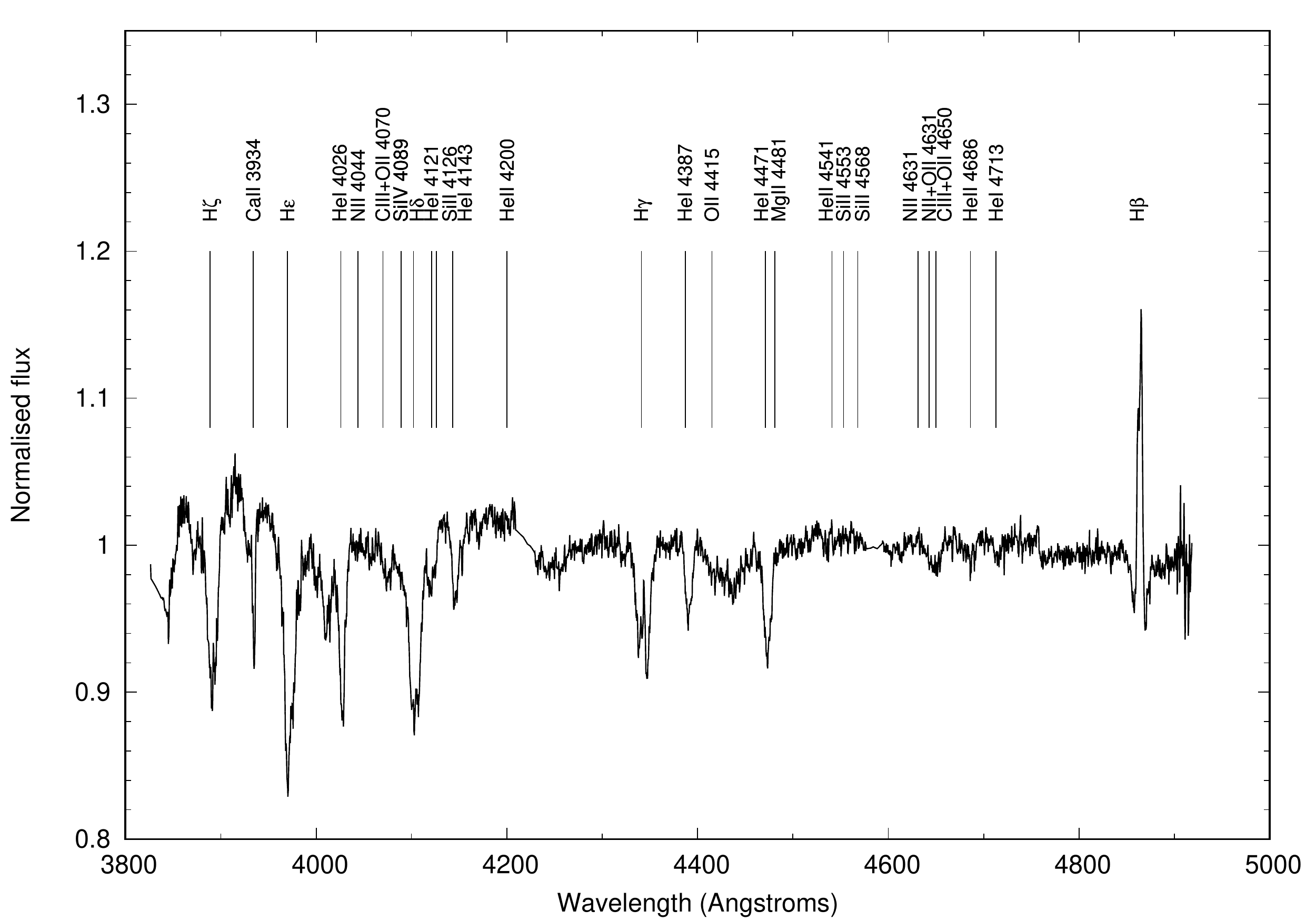}}
\caption{An example SALT blue spectrum of SXP4.78 with all the line species labelled.}
\label{fig:blue_spec}
\end{figure*}

We obtained optical spectra of SXP4.78 using the South African Large Telescope (SALT; \citealt{2006SPIE.6267E..0ZB}). The observations were performed using the Robert Stobie Spectrograph (RSS; \citealt{2003SPIE.4841.1463B,2003SPIE.4841.1634K}) in long slit mode between 16 November 2018 and 23 December 2018. Various grating set-ups were used, covering different wavelength regions: PG2300 ($3850-4900$~\AA), PG0900 ($4350-7400$~\AA) and PG1800 ($5985-7250$~\AA). Table~\ref{tab:observations_settings} summarises the set-up of the SALT observations. The primary reductions (overscan correction, bias subtraction, gain correction and amplifier cross-talk correction) were performed using the SALT pipeline \citep{2012ascl.soft07010C}. The remainder of the data reduction process, which consists of identification of the arc lines, background subtraction and extraction of the 1D spectra was undertaken using various tasks in \textsc{iraf}\footnote{Image Reduction and Analysis Facility: iraf.noao.edu}. Fig~\ref{fig:blue_spec} shows an example of the blue spectrum  with all the line species shown.

\begin{table}
	\centering
	\caption{A summary of the settings for the SALT observations}
	\label{tab:observations_settings}
    \setlength\tabcolsep{0.8pt}
	\begin{tabular}{ccccc} 
		\hline\hline
		 & Grating  & Wavelength & Exposure &    \\ [-3pt]
		Grating &  angle &  range (\AA) & time (s) & Resolution (\AA)  \\
		\hline
		PG0900 & 30.5 & $4350-7400$  & 1800 & 6.1 \\
		PG1800 & 36.875 & $5985-7250$  & 900 & 1.8 \\
		PG2300 & 30.5 & $3850-4900$  & 2000 & 2.1 \\
		\hline
	\end{tabular}
\end{table}

\subsection{OGLE}
\begin{figure}
\includegraphics[width=9cm,angle=0]{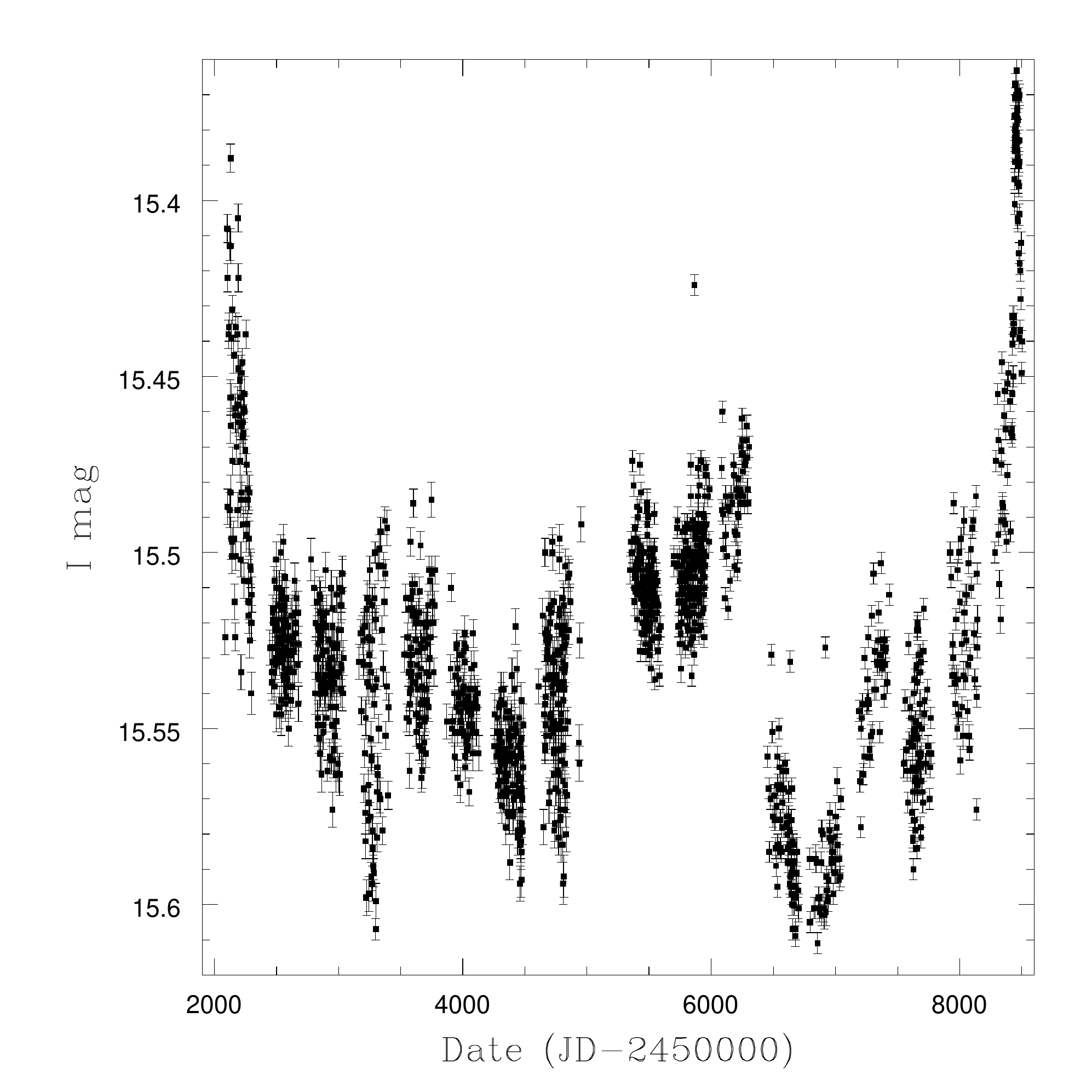}
\caption{The 17 years of OGLE $I$-band data from OGLE III and IV.}
\label{fig:ogle}
\end{figure}

The OGLE project \citep{1997AcA....47..319U, 2015AcA....65....1U}  provides long term I-band photometry with a cadence of 1-3 days. Once the optical counterpart of the source had been identified through the accurate Swift XRT position with the Be star [M2002] SMC~20671, it was possible to retrieve over 17 years of photometric monitoring in the I-band. The source is identified as SMC 719.21.22049 in OGLE IV and SMC 101.4.21932 in OGLE III. The whole lightcurve of the $I$-band magnitudes is shown in Fig.~\ref{fig:ogle} and reveals 2-3 major optical peaks over the period of the OGLE observations - the most recent being coincident with the current X-ray outburst. A detailed presentation of the current outburst may be seen in Fig.~\ref{fig:xi}.

\subsection{LCO}
CCD imaging photometry was obtained through the Las Cumbres Observatory (LCO) global network of robotically controlled telescopes, which currently consists of two 2m, nine 1m, and ten 0.4m instruments, distributed amongst six sites covering both hemispheres \citep{2013PASP..125.1031B,2014SPIE.9149E..1EB}. The optical imaging instrumentation (cameras and filters) on telescopes in each aperture class are identical, providing consistent datasets over long timescales (years). The data is all processed using the BANZAI automatic pipeline \citep{2018SPIE10707E..0KM}, which performs instrumental signature removal (bad pixel masking, bias and dark removal, flat-field correction), astrometric fitting and source catalogue extraction.
In this case observations were initially obtained from the 2m Faulkes Telescope South at Siding Spring, Australia, in the B (20~s exposure), V (10~s), R (10~s) and I (10~s) bands using the Spectral imaging system. Follow-up monitoring observations were then conducted using the LCO 1m facilities at Siding Spring and CTIO, Chile, in the SDSS g$^{\prime}$ (20--30~s), r$^{\prime}$ (10--20~s) and i$^{\prime}$ (10--20~s) bands using the Sinistro imaging system. The LCO observations for our programme span a period between 19 October 2018 and 13 December 2018.

\subsection{RXTE}
Once the association of XTE J0052--723 and Swift J005139--721704 had been confirmed, we searched the RXTE archive for historical detections of the source. We found no detections of the source in addition to the two outbursts published in the literature (\citealt{2003MNRAS.339..435L}; \citealt{2008ApJS..177..189G}. The discovery outburst peaked at roughly 10$^{38}$\,erg/s and lasted for at least 5 weeks \citep{2003MNRAS.339..435L}. The peak luminosity was derived from RXTE spectra, so should be used as a rough estimate only, though \citet{2003MNRAS.339..435L} made attempts to account for any non-pulsed emission from other sources in the field of view. RXTE measured the spin period of the system 5 times during the outburst, as shown in Table~\ref{tab:rxte} and originally listed in \citet{2003MNRAS.339..435L}.

\citet{2008ApJS..177..189G} show evidence for one other notable detection of the source around MJD 53725. This outburst lasted for around 50 days and had a similar peak luminosity to the discovery outburst. Some additional detections are noted in \citet{2014MNRAS.437.3863K}, but these are likely spurious events in the automated period detection routine used in the search for pulsars in RXTE data (see \citealt{2008ApJS..177..189G} for methods), as they are of lower significance and result in periods significantly different from the two bright archival outbursts and the current outburst.

There seems to be little evidence for periodic repetition in the three outbursts recorded for this source. There are roughly 1820 days and 4700 days between the first and second outburst and second and current outburst respectively. \citet{2017ApJ...839..119Y} report that the source was in the field of view of RXTE around 800 times, only being detected 11 times. The source was not detected by any other X-ray telescope until the current outburst, despite the deep XMM and Chandra surveys of the SMC.

\begin{table}
	\centering
	\caption{RXTE spin period history, except the last value which is from the 2018-19 outburst obtained with NuSTAR \citep{2003MNRAS.339..435L,2018ATel12234....1A}}
	\label{tab:rxte}
    \setlength\tabcolsep{2pt}
	\begin{tabular}{ccc} 
		\hline\hline
		MJD & Spin period (s) & Error (s) \\
		\hline
		51905  &  4.7817   &   0.0001  \\
        51912  &  4.7818   &   0.0001  \\
        51918  &  4.782    &   0.001   \\
        51927  &  4.7820   &   0.0001  \\
        51933  &  4.7823   &   0.0001  \\
        53725  &  4.78126  &   0.00009 \\
        53733  &  4.78149  &   0.00015 \\
        53738  &  4.78149  &   0.00007 \\
        53747  &  4.78172  &   0.00007 \\
        53754  &  4.78172  &   0.00012 \\
        53774  &  4.78172  &   0.00013 \\
        58439  &  4.781607  &   0.000007 \\
		\hline
	\end{tabular}
\end{table}

\begin{figure}
\includegraphics[width=8cm,angle=0]{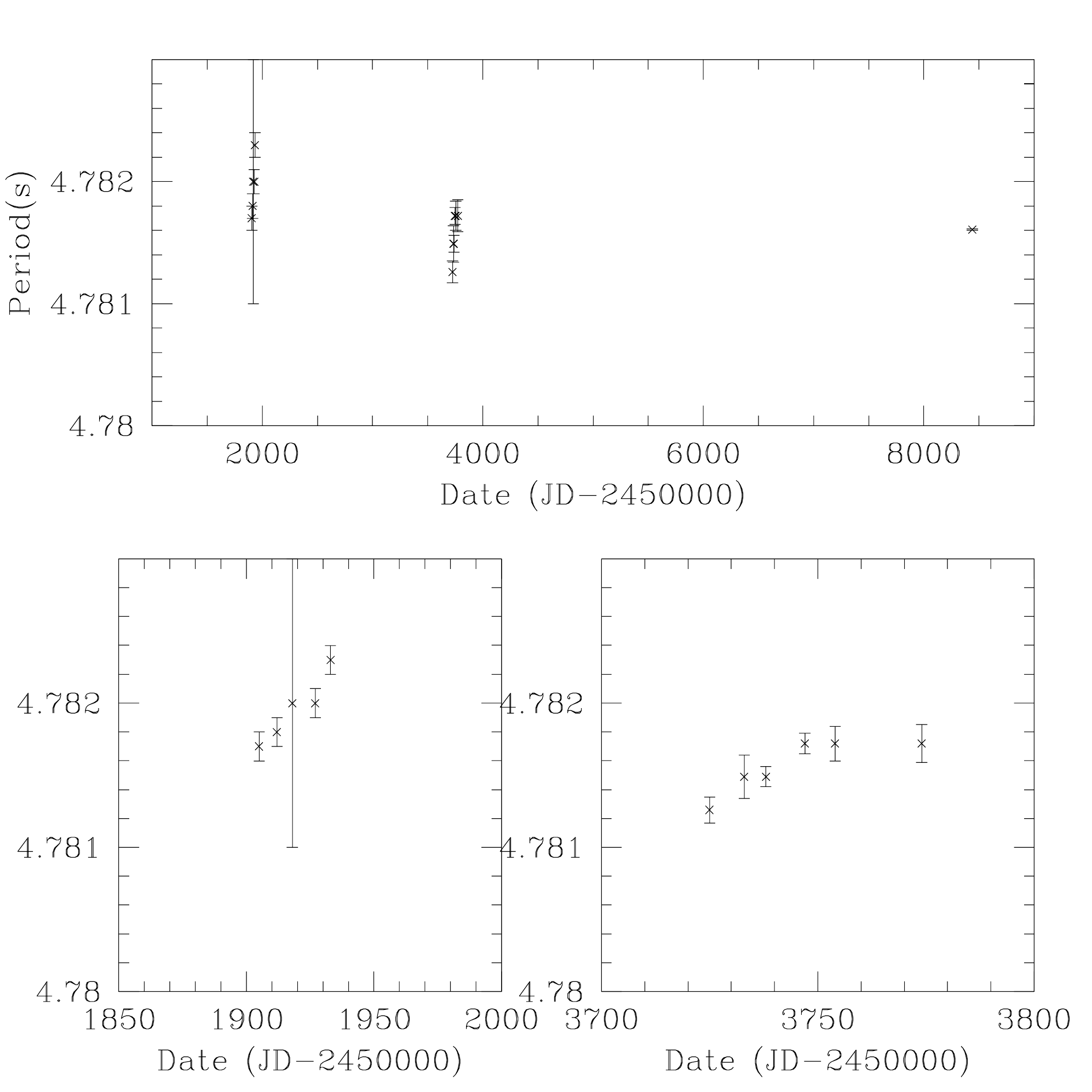}
\caption{(Top panel) Published spin period history of SXP4.78. All points except the last one are from RXTE; the last one comes from NICER - see text for references. (Lower 2 panels) The details of the spin period evolution as seen by RXTE during two previous outbursts.}
\label{fig:rxte-pulse}
\end{figure}

\section{Results \& Discussion}
\label{sec:results}

\subsection{Search for the binary period}

\subsubsection{OGLE data}
\label{sec:ogle}
It is well-known that many Be/X-ray binary systems reveal evidence for a binary modulation in the OGLE $I$-band data \citep{2012MNRAS.423.3663B}. It is believed that this optical signature arises from the disruption of the circumstellar disk close to the time of the periastron passage of the neutron star. 

Since SXP 4.78 has been monitored by the OGLE project for over 17 years (see Fig.~\ref{fig:ogle}), we searched these data for a binary signature. It is clear from that figure that the $I$-band has peaked 2-3 times over this time frame, the most recent peak correlating with the X-ray outburst discussed in this paper.

The Corbet diagram \citep{1984A&A...141...91C}  shows a general correlation between pulse and binary periods and hence provides a useful guide to the most probable time domain to search. In the case of SXP 4.78 it is the range is 20 - 50 days, though the search was extended to cover all periods from 2d to 300d. The data were de-trended using a 3rd order polynomial and a Lomb-Scargle routine was used for the search, but no significant peak was found at any time over the whole search range. In addition, smaller subsets of the OGLE data were studied (for example the most recent 2-3 years), but no evidence for a significant coherent periodic signature was found. The largest peak seen at any time corresponded to a period of 2.65d but that is most likely related to non-radial pulsation (NRP) behaviour in the Be star. Certainly when the data are folded at that period the resulting profile is sinusoidal in nature, consistent with such an explanation. 

\subsubsection{RXTE data}

The remarkable stability of the pulse period of SXP4.78 is evident in Table \ref{tab:rxte} and Fig. \ref{fig:rxte-pulse}, and continues to be so during the current outburst \citep{2018ATel12234....1A}. It is clear from Fig. \ref{fig:ogle} that the discovery outburst coincided with a previous period of enhanced disk emission (the only other time the optical flux of this source has been recorded above 15.4 magnitudes in the \textit{I}-band). Interestingly, the second X-ray outburst around MJD 53725 coincides with a period of low optical flux and hence likely a smaller disk. This is yet another example of X-ray outbursts occurring in Be/X-ray binaries when the circumstellar disk appears to be small. In other instances where such behaviour was observed the implication of the small disc size (traced through the H$\alpha$ EW) was suggested to be due to a precessing warped disc, where the low values of the H$\alpha$ EW are due to geometric effects \citep{2013PASJ...65...83M}. Another possible reason for this could be the presence of an eccentric disc with the elongation part pointing away from the observer \citep{2014ApJ...790L..34M,2017MNRAS.464..572M}.   

It is unusual to observe a NS spinning down during a period of enhanced accretion. Fig. \ref{fig:rxte-pulse} clearly shows this is the case for both previous outbursts of this source. It is difficult to say from these data whether the cause of the spin-down is due to accretion torques (in which case the torque would be acting in the opposite direction to the spin of the NS), or whether it is due to Doppler shifting. If the latter, we can constrain the orbital period to greater than roughly 100 days as this is twice the length of the longest outburst, in which we only see the spin period increase. Thus we are not seeing the corresponding decrease from motion on the other side of the orbit. This is slightly longer than suggested by the Corbet diagram, but certainly possible. The period changes are likely due to a combination of both factors, making constraining the orbital period difficult.

\subsubsection{Radial velocity search}
We performed radial velocity (RV) analysis using the blue spectra taken with the PG2300 grating. This was done using cross-correlation, by applying the iterative procedure described in \citet{2003MNRAS.338..360F}, \citet{2015MNRAS.448.1789M} and \citet{2017ApJ...847...68M} to create a high signal-to-noise template, which is summarised as follows:
\begin{itemize}
    \item After the spectra were corrected for the heliocenter, they were  normalised and then the continuum level was subtracted.
    \item The spectra were then converted to a logarithmic wavelength scale and then sorted on the basis of their signal-to-noise ratio.
    \item The spectrum with the highest signal-to-noise ratio was then used as template in the first iteration of cross-correlation. 
    \item The spectra were then shifted to the same wavelength as the template spectrum using the results of the first cross-correlation iteration.
    \item The spectra were then combined to create a high signal-to-noise spectrum to be used as a template for the final cross-correlation iteration.
\end{itemize}

\begin{figure}
\resizebox{\hsize}{!}
            {\includegraphics[width=10cm,angle=0]{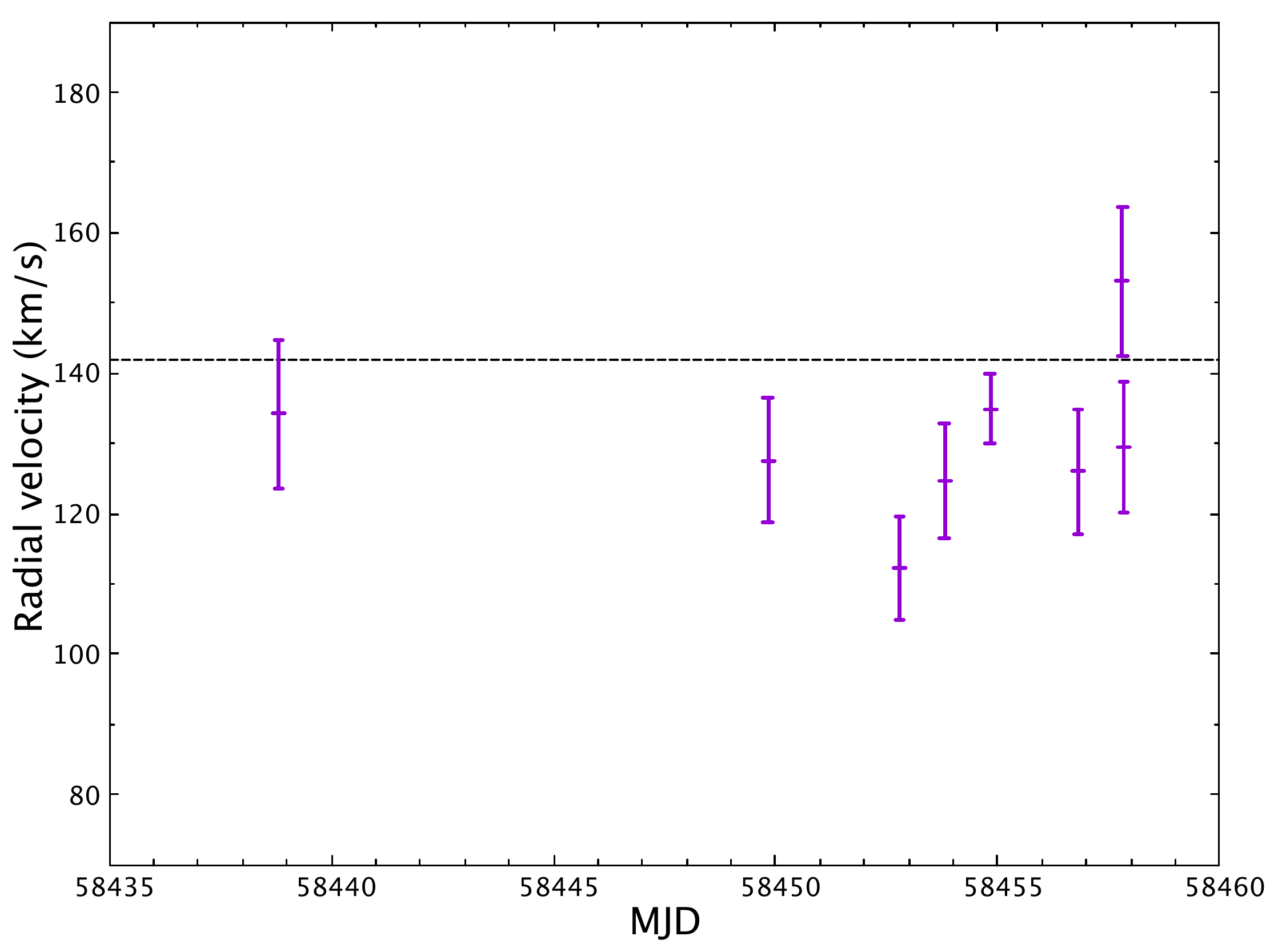}}
\caption{The measured radial velocities of SXP4.78. The dashed line shows the recessional velocity of the SMC.}
\label{fig:RV_variability}
\end{figure}

Table~\ref{tab:RV_values} is a summary of the RV shifts. Due to the infilling present in the Balmer lines, we performed cross-correlation analysis considering only the metal lines. Different combinations of these line species were considered. The analysis was also performed using all the lines available in the blue spectra with little differences in the overall variability. Fig.~\ref{fig:RV_variability} shows the RV measurements from the whole wavelength range. The measured RVs shown here display little evidence of any periodicity, but rather show an average velocity consistent with that of the redshift velocity of the SMC \citep{2012AJ....144....4M}. We note that the semi-amplitude is expected to be very low (on the order of 20 km/s), considering the spectral type of the primary (see section~\ref{sec:spec_type}) and a lower limit of the orbital period of 20 days, making it difficult to detect any periodicity with our poor sampling and relatively large error bars of the RVs. 

\begin{table}
	\centering
	\caption{A summary of the RV measurements of SXP4.78}
	\label{tab:RV_values}
    \setlength\tabcolsep{2pt}
	\begin{tabular}{cc} 
		\hline\hline
		MJD & Radial velocity (km~s$^{-1}$)  \\
		\hline
		58438.8267593  &   144  $\pm$ 11 \\
        58449.8244676  &   137.7  $\pm$  8.9 \\
        58452.8029398  &   122.3  $\pm$  7.4 \\
        58453.7979977  &   134.7  $\pm$  8.1 \\
        58454.8230093  &   144.9  $\pm$  4.9 \\
        58456.8003588  &   136.0  $\pm$  8.9 \\
        58457.8088773  &   163  $\pm$  11 \\
        58457.8325231  &   139.6  $\pm$  9.3 \\
		\hline
	\end{tabular}
\end{table}

\subsection{Spectral classification}
\label{sec:spec_type}
The blue spectrum in Fig.~\ref{fig:blue_spec} shows the sum of 8 observations with a combined averaged exposure of 1800~s. The spectrum is clearly an early type star with many of the Balmer lines showing infilling, likely from the circumstellar disc emission.  The C\rom{3}+O\rom{2} blend at 4650 constrains the spectral type to earlier than B3 \citep{2004PASP..116..909E} and the weak He II 4686 further constrains the spectral type to B0.5. Most of the luminosity indicators at this spectral type involve metallic lines species, so we use the observed magnitude ($V-15.59$) together with the distance modulus of the SMC (18.95, \citealt{2013IAUS..289..222G}) to constrain the luminosity class to a \rom{4}-\rom{5} star. The spectral type is slightly earlier that that of both Maravelias \citep{2018ATel12224....1M} and Evans \citep{2004PASP..116..909E} due to our significantly increased exposure. This enabled detection of a very faint He\rom{2} 4686 line.

\subsection{The CMD}
\begin{table}
	\centering
	\caption{Magnitude range for the different filters obtained from observations reported in this work.}
	\label{tab:optical_mags}
    \setlength\tabcolsep{2pt}
	\begin{tabular}{cc} 
		\hline\hline
		Filter & magnitude range  \\
		\hline
 $I$ & 15.6 $-$ 15.3 \\
 $V$ & 15.7 $-$ 15.5 \\
 $B$ & 15.8 $-$ 15.6\\
 $g^{'}$ &  15.6 $-$ 15.5\\
 $r^{'}$ & 15.6 $-$ 15.7 \\ 
 $i^{'}$ & 15.8 $-$ 15.7\\
		\hline
	\end{tabular}
\end{table}

\begin{figure}
\resizebox{\hsize}{!}
            {\includegraphics[angle=0]{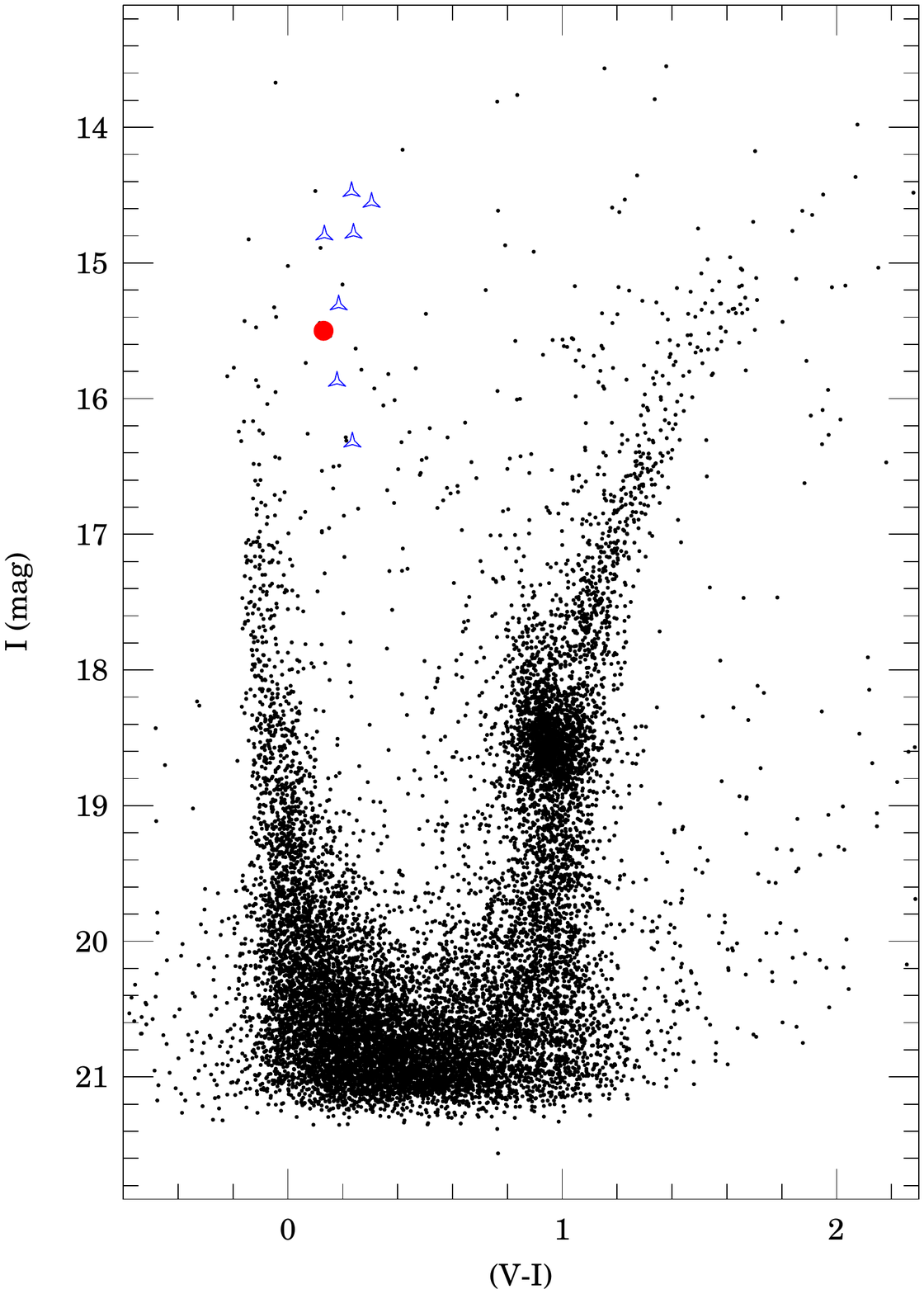}}
\caption{The location of SXP4.78 on an OGLE CMD. The location of SXP 4.78 is shown by the red circle. Also shown are the location of seven of the Be/X-ray binary systems in the SMC by blue triangles (see text for list).}
\label{fig:cmd}
\end{figure}

A colour-magnitude diagram (CMD)  of the 8.6 x 8.6 arcmin region around SXP 4.78 was produced based on the OGLE-III photometric maps of the SMC \citep{2008AcA....58..329U}. The result is shown is shown in Fig.~\ref{fig:cmd}. The red dot indicates the position of SXP 4.78. It is worth noting that though the star is not located on the SMC main sequence its position is consistent with other Be/X-ray binary systems in the SMC. Seven such sources were chosen at random and are indicated by blue triangles. The comparison sources used were SXP 7.78, SXP 9.13, SXP 15.3, SXP 31.0, SXP 91.1, SXP 264 and SXP 293. Details of these comparison sources may be found in \cite{2015MNRAS.452..969C}. Assuming a distance modulus of 18.95 to the SMC \citep{2013IAUS..289..222G} and a reddening of E(B-V) = 0.08 \citep{1991A&A...246..231S} a B1V star should exhibit an observed  (V-I) colour of -0.018. The colour reported here is (V-I) $\sim$+0.1 indicating considerable local reddening. It is highly likely that in all SXP systems the excess reddening is caused by local absorption associated with the Be star's circumstellar disk and wind outflow.

\subsection{Optical colour changes}

\begin{figure}
\resizebox{\hsize}{!}
            {\includegraphics[angle=0]{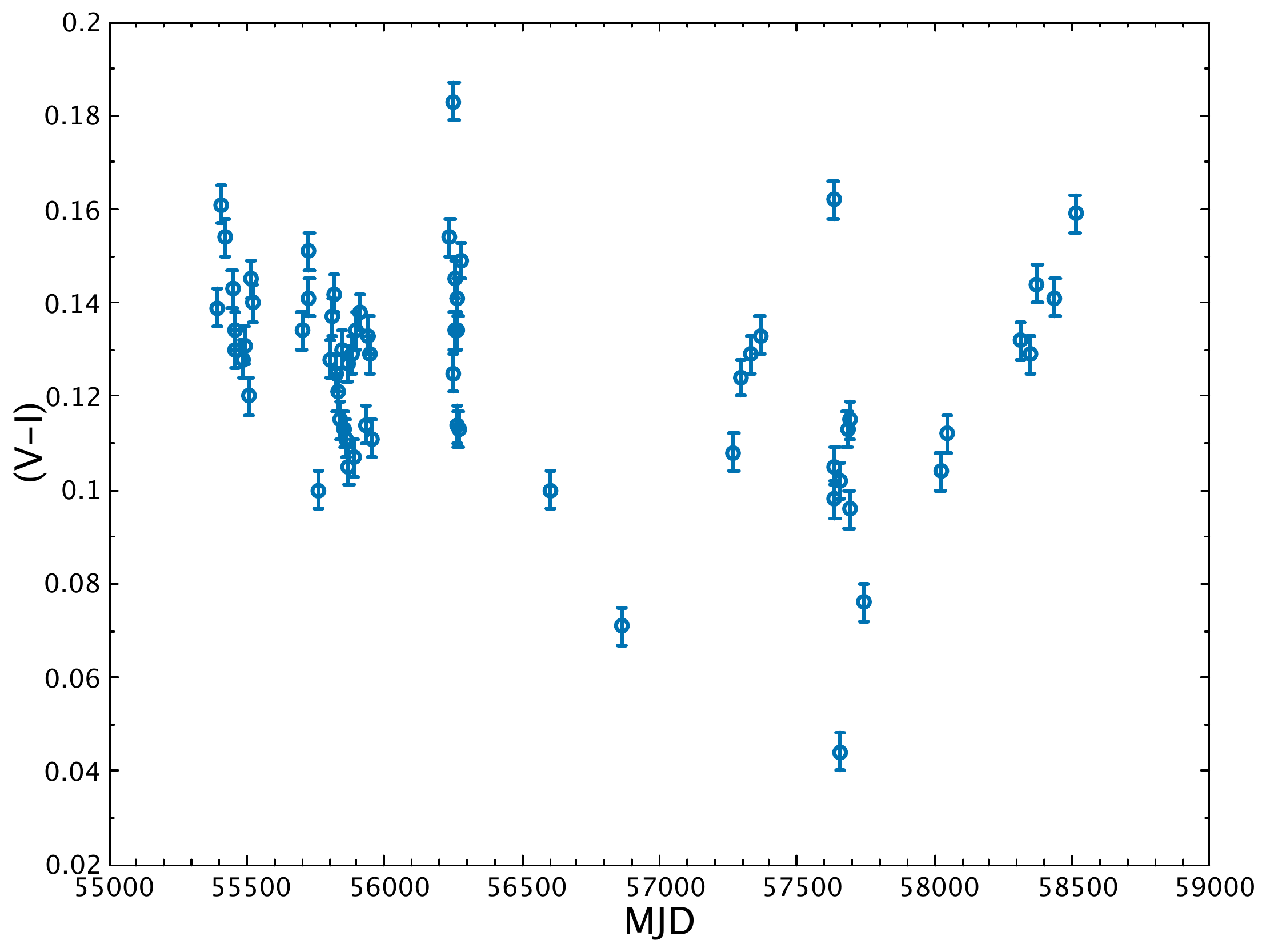}}
\caption{Long-term evolution of the $(V-I)$ colour index - data from the OGLE project.}
\label{fig:VI_MJD}
\end{figure}

\begin{figure}
\resizebox{\hsize}{!}
            {\includegraphics[angle=0]{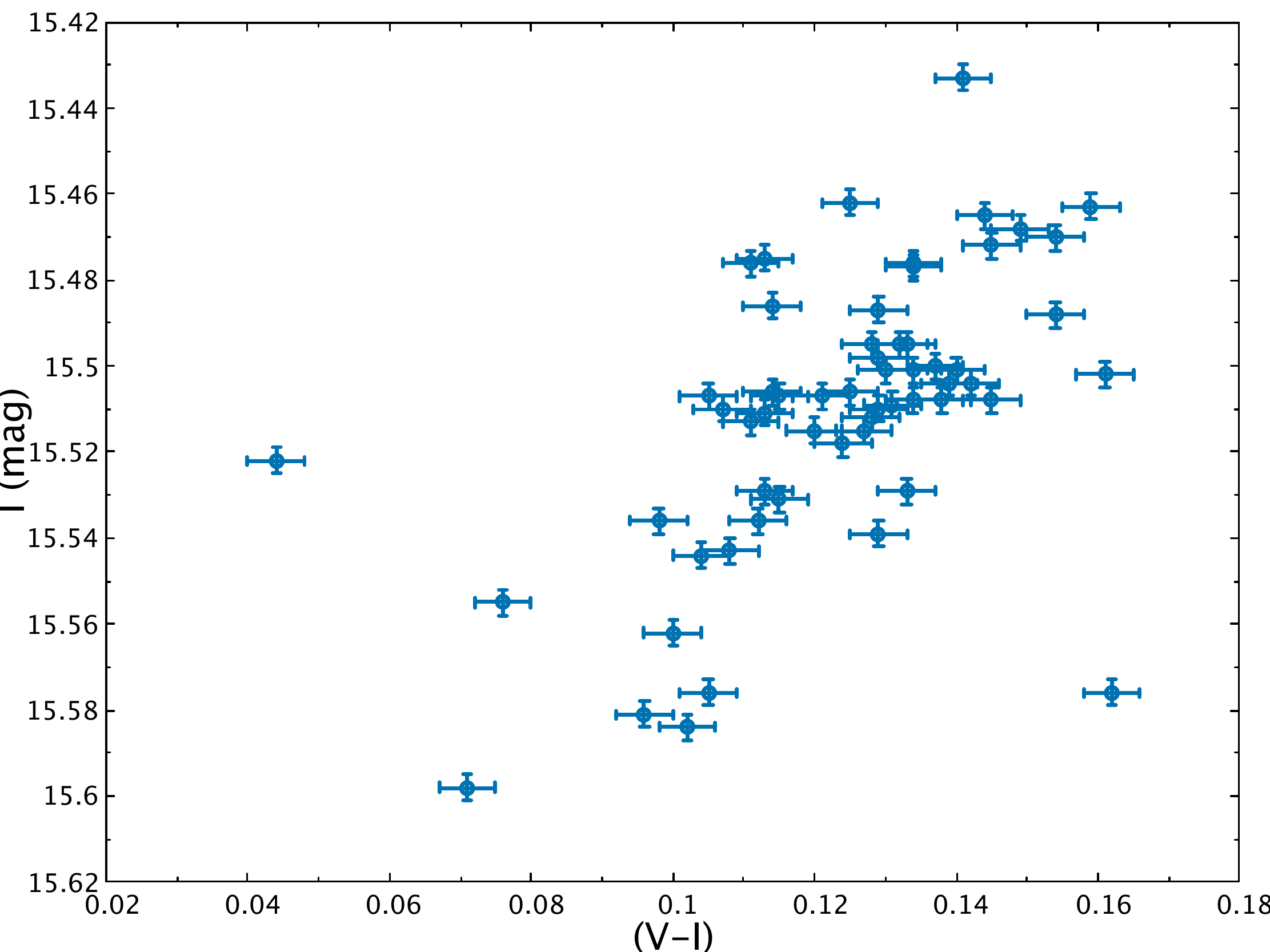}}
\caption{$(V-I)$ - $I$ colour magnitude diagram.}
\label{fig:VI_I}
\end{figure}

\begin{figure}
\resizebox{\hsize}{!}
            {\includegraphics[angle=0]{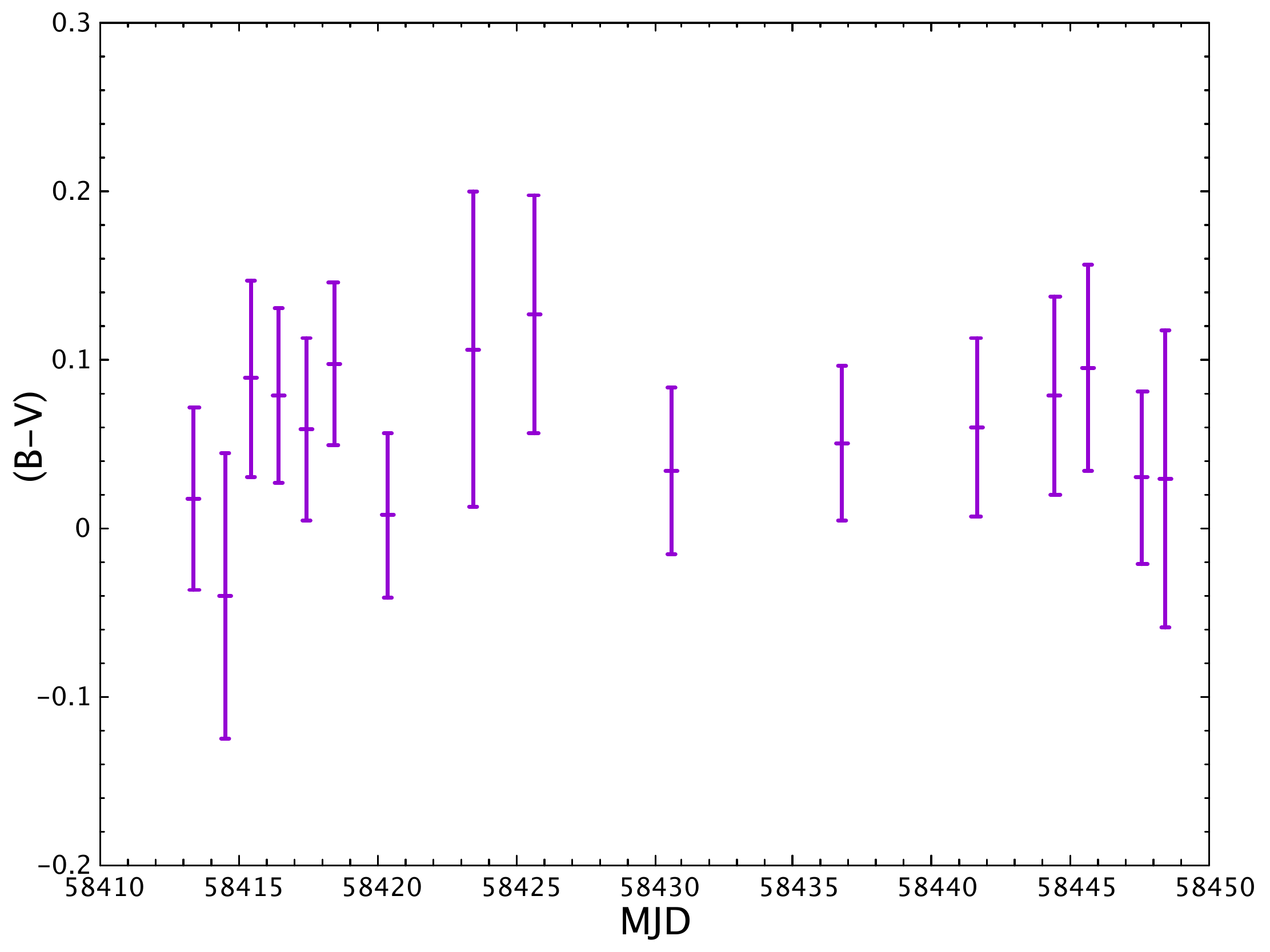}}
\caption{Evolution of the $(B-V)$ colour index during the recent Type~\rom{2} outburst. Data from LCO observations.}
\label{fig:BV}
\end{figure}

Figure~\ref{fig:VI_MJD} shows the long-term evolution of the $(V-I)$ colour index from OGLE, where we see relatively large amplitude changes attributed to the Be disc variability. Figure~\ref{fig:VI_I} shows the $(V-I)$ - $I$ colour magnitude diagram, where a positive correlation is seen, i.e. the system reddens as it increases in brightness. Such behaviour has been previously observed in several systems where the variability is explained by a geometric effect resulting from a low viewing angle of the Be disc relative to the observer which causes an increase in brightness and red continuum as the radius of the disc increases \citep{1983HvaOB...7...55H,2011MNRAS.413.1600R,2015A&A...574A..33R}. The suggestion of a low inclination of the Be disc is corroborated by the H$\alpha$ emission line having a single-peaked morphology in all our observations (see section~\ref{sec:Be_disc}). Figure~\ref{fig:BV} displays the LCO $(B-V)$ colour index for observations taken during the rise of the recent outburst where no significant variability in the short-term measurements is seen.

\subsection{Be disc variability}
\label{sec:Be_disc}
\begin{figure}
\resizebox{\hsize}{!}
            {\includegraphics[width=10cm,angle=0]{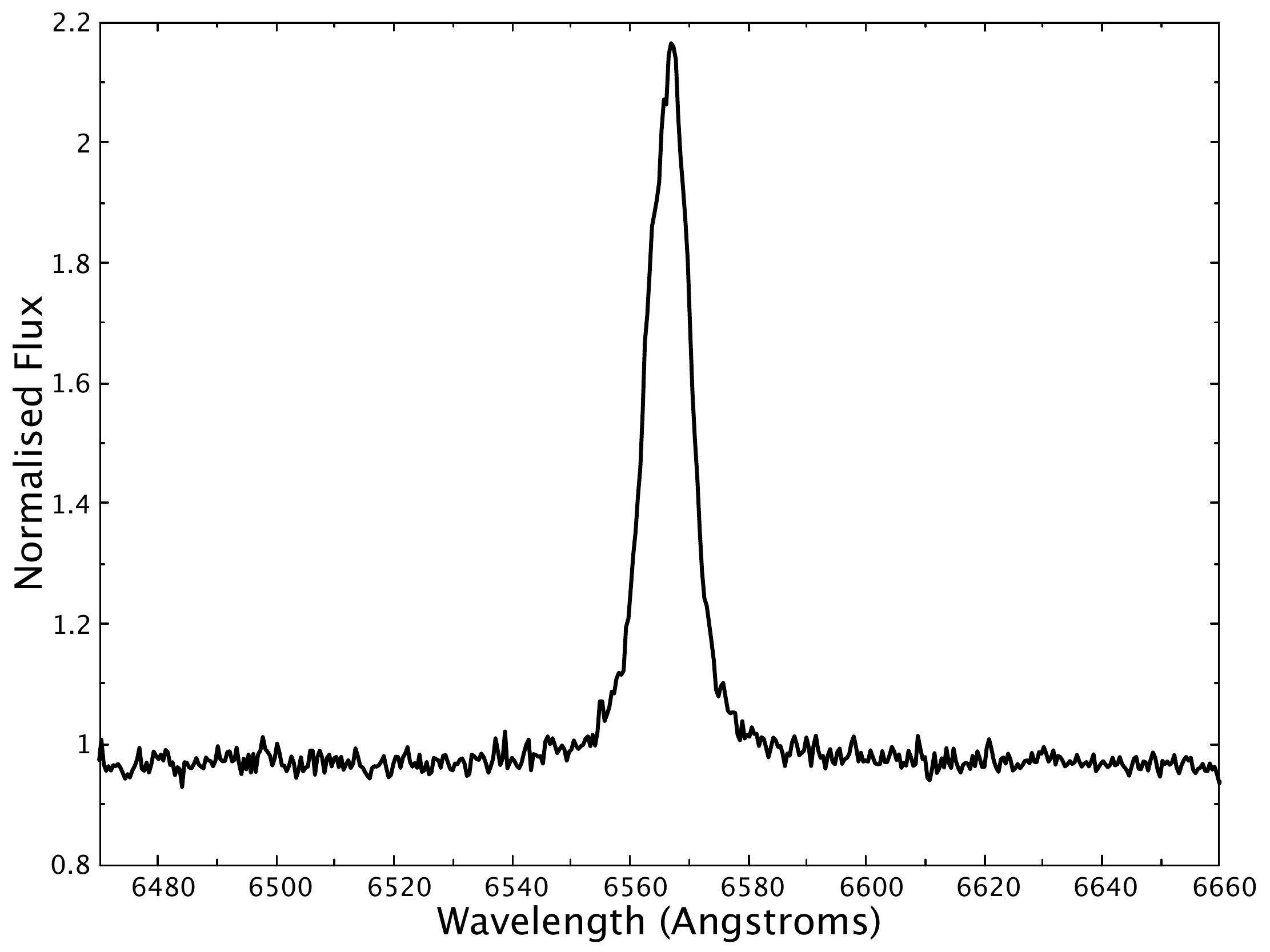}}
\caption{An example of the H$\alpha$ emission line in SXP4.78}
\label{fig:Halpha}
\end{figure}

\begin{figure}
\resizebox{\hsize}{!}
            {\includegraphics[width=10cm,angle=0]{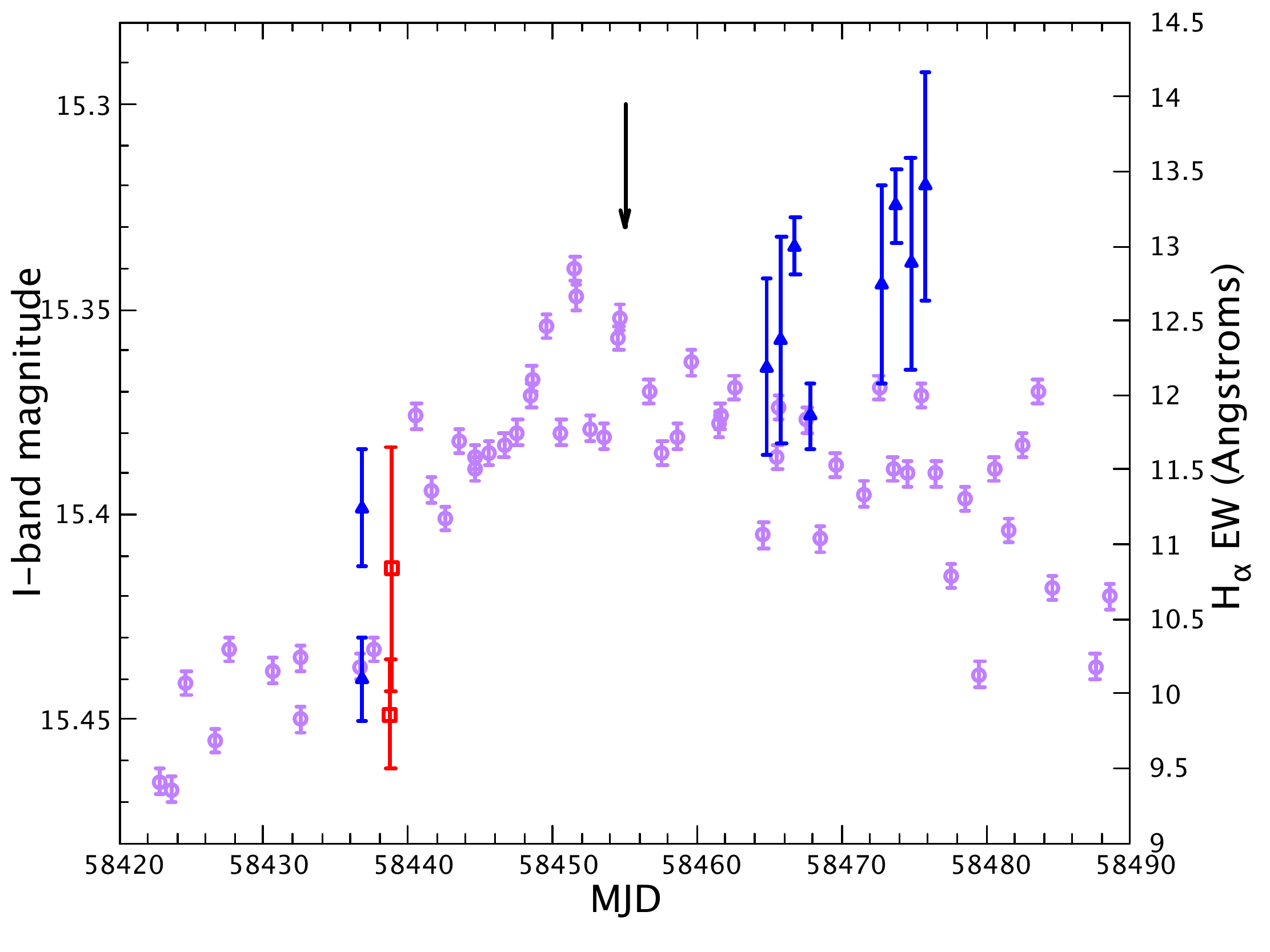}}
\caption{OGLE $I$-band light curve (purple circles) and H$\alpha$ EW  measurements from PG900 (red squares) and PG1800 (blue triangles) gratings. The arrow indicates the epoch of the X-ray outburst peak.}
\label{fig:OGLE_EW}
\end{figure}

In addition to the blue spectra, we have obtained spectra covering the red end. The H$\alpha$ line is present in emission and shows a single-peaked profile in all our spectra (Fig.~\ref{fig:Halpha} is an example of the H$\alpha$ emission line profile), suggesting the presence of the circumstellar disc throughout our period of monitoring. The H$\alpha$ equivalent width (EW) is used as an indicator of the size of the disc in Be stars. Fig.~\ref{fig:OGLE_EW} shows the evolution of the EW, where the recent measurements (58466 < MJD < 58475) have shown a slight increase from those at the start of the monitoring campaign (58436 < MJD < 58438), suggesting disc growth in this period. The measured EW values are given in Table~\ref{tab:EW_measurements}. Fig.~\ref{fig:OGLE_EW} also shows the OGLE $I$-band lightcurve overplotted with the EW measurements, and we note that the gradual increase in the $I$-band flux until it reaches a high state matches the EW behaviour, further supporting the suggestion of disc growth. We note that \citet{2004PASP..116..909E} report the H$\alpha$ line in absorption from previous observations. \citet{1997A&A...322..193R} reported the existence of a correlation between the maximum H$\alpha$ EW and orbital period in BeXBs. From a rough interpolation from that relationship and using the maximum value of the H$\alpha$ EW measured in this work (13.40 \AA), this results in an orbital period of $\sim$25~days. This is in good agreement with the range of orbital periods suggested based on the Corbet diagram (see section~\ref{sec:ogle}).

\begin{table}
	\centering
	\caption{H$\alpha$ equivalent width measurements of SXP4.78}
	\label{tab:EW_measurements}
    \setlength\tabcolsep{2pt}
	\begin{tabular}{ccc} 
		\hline\hline
		MJD & EW (\AA) & Grating  \\
		\hline
58436.7981482 & 10.09  $\pm$ 0.28 & PG1800 \\
58436.7945139 & 11.24 $\pm$  0.39 & PG1800 \\
58438.8108333 & 10.84 $\pm$  0.82 & PG900\\
58438.8002662 & 9.86 $\pm$  0.37 & PG900\\
58466.8024421 & 13.00 $\pm$  0.19 & PG1800\\
58464.8145486 & 12.19 $\pm$  0.59 & PG1800\\
58465.8335648 & 12.37 $\pm$  0.70 & PG1800\\
58467.8104398 & 11.86  $\pm$ 0.21 & PG1800\\
58436.7981482 & 10.09 $\pm$  0.28 & PG1800\\
58436.7945139 & 11.24 $\pm$  0.39 & PG1800\\
58472.8146412 & 12.74 $\pm$  0.66 & PG1800\\
58473.7998611 & 13.27 $\pm$  0.24 & PG1800\\
58474.7845833 & 12.88 $\pm$  0.72 & PG1800\\
58475.7893171 & 13.40 $\pm$  0.77 & PG1800\\
		\hline
	\end{tabular}
\end{table}

\section{Conclusion}
In this work we have presented X-ray and optical analysis of SXP4.78 studying its historic and recent behaviour. We have shown that the Be disc has undergone long-term variability through the analysis of photometric flux and colour changes. We conducted period searches using both photometric and spectroscopic data with no significant detections of any orbital modulation. A detection of a period at 2.65~days was made from the long-term OGLE $I$--band data which is most likely associated with the NRPs of the Be star. The orbital period remains elusive at this time. We have shown that the circumstellar disc has exhibited recent growth through the increase in optical magnitude and H$\alpha$ EW. We believe that it is this disc growth that has supplied matter resulting in the Type~\rom{2} outburst. Our blue spectrum from SALT permits us to classify the spectral type as B0.5 IV-V, typical for an optical counterpart to a Be/X-ray binary system.

\section*{Acknowledgements}
The OGLE project has received funding from the National Science Centre,
Poland, grant MAESTRO 2014/14/A/ST9/00121 to AU. DAHB, LJT and IMM are supported by the South African NRF. Some of the observations reported in this paper were obtained with the Southern African Large Telescope (SALT), as part of the Large Science Programme on transients 2018-2-LSP-001 (PI: Buckley). Polish participation in SALT is funded by grant no. MNiSW DIR/WK/2016/07. This work makes use of observations from the LCOGT network. This work is based on the research supported by the National Research Foundation of South Africa (Grant numbers 98969 and 93405).

\label{lastpage}

\bibliographystyle{mnras}
\bibliography{references}

\begin{thebibliography}{}
\makeatletter
\relax
\def\mn@urlcharsother{\let\do\@makeother \do\$\do\&\do\#\do\^\do\_\do\%\do\~}
\def\mn@doi{\begingroup\mn@urlcharsother \@ifnextchar [ {\mn@doi@}
  {\mn@doi@[]}}
\def\mn@doi@[#1]#2{\def\@tempa{#1}\ifx\@tempa\@empty \href
  {http://dx.doi.org/#2} {doi:#2}\else \href {http://dx.doi.org/#2} {#1}\fi
  \endgroup}
\def\mn@eprint#1#2{\mn@eprint@#1:#2::\@nil}
\def\mn@eprint@arXiv#1{\href {http://arxiv.org/abs/#1} {{\tt arXiv:#1}}}
\def\mn@eprint@dblp#1{\href {http://dblp.uni-trier.de/rec/bibtex/#1.xml}
  {dblp:#1}}
\def\mn@eprint@#1:#2:#3:#4\@nil{\def\@tempa {#1}\def\@tempb {#2}\def\@tempc
  {#3}\ifx \@tempc \@empty \let \@tempc \@tempb \let \@tempb \@tempa \fi \ifx
  \@tempb \@empty \def\@tempb {arXiv}\fi \@ifundefined
  {mn@eprint@\@tempb}{\@tempb:\@tempc}{\expandafter \expandafter \csname
  mn@eprint@\@tempb\endcsname \expandafter{\@tempc}}}

\bibitem[\protect\citeauthoryear{{Antoniou}, {Zezas}, {Hong}, {Kennea},
  {Tomsick}  \& {Haberl}}{{Antoniou} et~al.}{2018}]{2018ATel12234....1A}
{Antoniou} V.,  {Zezas} A.,  {Hong} J.,  {Kennea} J.,  {Tomsick} J.,   {Haberl}
  F.,  2018, The Astronomer's Telegram, \href
  {http://adsabs.harvard.edu/abs/2018ATel12234....1A} {12234}

\bibitem[\protect\citeauthoryear{{Bird}, {Coe}, {McBride}  \& {Udalski}}{{Bird}
  et~al.}{2012}]{2012MNRAS.423.3663B}
{Bird} A.~J.,  {Coe} M.~J.,  {McBride} V.~A.,   {Udalski} A.,  2012, \mn@doi
  [\mnras] {10.1111/j.1365-2966.2012.21163.x}, \href
  {http://ukads.nottingham.ac.uk/abs/2012MNRAS.423.3663B} {423, 3663}

\bibitem[\protect\citeauthoryear{{Boroson} et~al.,}{{Boroson}
  et~al.}{2014}]{2014SPIE.9149E..1EB}
{Boroson} T.,  et~al., 2014, in Observatory Operations: Strategies, Processes,
  and Systems V. p. 91491E, \mn@doi{10.1117/12.2054776}

\bibitem[\protect\citeauthoryear{{Brown} et~al.,}{{Brown}
  et~al.}{2013}]{2013PASP..125.1031B}
{Brown} T.~M.,  et~al., 2013, \mn@doi [\pasp] {10.1086/673168}, \href
  {http://adsabs.harvard.edu/abs/2013PASP..125.1031B} {125, 1031}

\bibitem[\protect\citeauthoryear{{Buckley}, {Swart}  \& {Meiring}}{{Buckley}
  et~al.}{2006}]{2006SPIE.6267E..0ZB}
{Buckley} D.~A.~H.,  {Swart} G.~P.,   {Meiring} J.~G.,  2006, in Society of
  Photo-Optical Instrumentation Engineers (SPIE) Conference Series. p. 62670Z,
  \mn@doi{10.1117/12.673750}

\bibitem[\protect\citeauthoryear{{Burgh}, {Nordsieck}, {Kobulnicky},
  {Williams}, {O'Donoghue}, {Smith}  \& {Percival}}{{Burgh}
  et~al.}{2003}]{2003SPIE.4841.1463B}
{Burgh} E.~B.,  {Nordsieck} K.~H.,  {Kobulnicky} H.~A.,  {Williams} T.~B.,
  {O'Donoghue} D.,  {Smith} M.~P.,   {Percival} J.~W.,  2003, in {Iye} M.,
  {Moorwood} A.~F.~M.,  eds,  \procspie Vol. 4841, Instrument Design and
  Performance for Optical/Infrared Ground-based Telescopes. pp 1463--1471,
  \mn@doi{10.1117/12.460312}

\bibitem[\protect\citeauthoryear{{Casares}, {Negueruela}, {Rib{\'o}}, {Ribas},
  {Paredes}, {Herrero}  \& {Sim{\'o}n-D{\'{\i}}az}}{{Casares}
  et~al.}{2014}]{2014Natur.505..378C}
{Casares} J.,  {Negueruela} I.,  {Rib{\'o}} M.,  {Ribas} I.,  {Paredes} J.~M.,
  {Herrero} A.,   {Sim{\'o}n-D{\'{\i}}az} S.,  2014, \mn@doi [\nat]
  {10.1038/nature12916}, \href
  {http://adsabs.harvard.edu/abs/2014Natur.505..378C} {505, 378}

\bibitem[\protect\citeauthoryear{{Coe} \& {Kirk}}{{Coe} \&
  {Kirk}}{2015}]{2015MNRAS.452..969C}
{Coe} M.~J.,  {Kirk} J.,  2015, \mn@doi [\mnras] {10.1093/mnras/stv1283}, \href
  {http://adsabs.harvard.edu/abs/2015MNRAS.452..969C} {452, 969}

\bibitem[\protect\citeauthoryear{{Coe}, {Kennea}, {Buckley}  \&
  {Udalski}}{{Coe} et~al.}{2018a}]{2018ATel12209....1C}
{Coe} M.~J.,  {Kennea} J.~A.,  {Buckley} D.,   {Udalski} A.,  2018a, The
  Astronomer's Telegram, \href
  {http://adsabs.harvard.edu/abs/2018ATel12209....1C} {12209}

\bibitem[\protect\citeauthoryear{{Coe}, {Kennea}, {Buckley}, {McBride},
  {Udalski}, {Evans}, {Roche}  \& {Townsend}}{{Coe}
  et~al.}{2018b}]{2018ATel12229....1C}
{Coe} M.~J.,  {Kennea} J.~A.,  {Buckley} D.,  {McBride} V.,  {Udalski} A.,
  {Evans} P.,  {Roche} P.,   {Townsend} L.,  2018b, The Astronomer's Telegram,
  \href {http://adsabs.harvard.edu/abs/2018ATel12229....1C} {12229}

\bibitem[\protect\citeauthoryear{{Coleiro} \& {Chaty}}{{Coleiro} \&
  {Chaty}}{2013}]{2013ApJ...764..185C}
{Coleiro} A.,  {Chaty} S.,  2013, \mn@doi [\apj] {10.1088/0004-637X/764/2/185},
  \href {http://adsabs.harvard.edu/abs/2013ApJ...764..185C} {764, 185}

\bibitem[\protect\citeauthoryear{{Corbet}}{{Corbet}}{1984}]{1984A&A...141...91C}
{Corbet} R.~H.~D.,  1984, \aap, \href
  {http://ukads.nottingham.ac.uk/abs/1984A%26A...141...91C} {141, 91}

\bibitem[\protect\citeauthoryear{{Corbet}, {Marshall}  \& {Markwardt}}{{Corbet}
  et~al.}{2001}]{2001IAUC.7562....1C}
{Corbet} R.,  {Marshall} F.~E.,   {Markwardt} C.~B.,  2001, \iaucirc, \href
  {http://adsabs.harvard.edu/abs/2001IAUC.7562....1C} {7562}

\bibitem[\protect\citeauthoryear{{Crawford} et~al.,}{{Crawford}
  et~al.}{2012}]{2012ascl.soft07010C}
{Crawford} S.~M.,  et~al., 2012, {PySALT: SALT science pipeline}, Astrophysics
  Source Code Library (\mn@eprint {ascl} {1207.010})

\bibitem[\protect\citeauthoryear{{Evans}, {Lennon}, {Walborn}, {Trundle}  \&
  {Rix}}{{Evans} et~al.}{2004}]{2004PASP..116..909E}
{Evans} C.~J.,  {Lennon} D.~J.,  {Walborn} N.~R.,  {Trundle} C.,   {Rix} S.~A.,
   2004, \mn@doi [\pasp] {10.1086/425563}, \href
  {http://adsabs.harvard.edu/abs/2004PASP..116..909E} {116, 909}

\bibitem[\protect\citeauthoryear{{Foellmi}, {Moffat}  \& {Guerrero}}{{Foellmi}
  et~al.}{2003}]{2003MNRAS.338..360F}
{Foellmi} C.,  {Moffat} A.~F.~J.,   {Guerrero} M.~A.,  2003, \mn@doi [\mnras]
  {10.1046/j.1365-8711.2003.06052.x}, \href
  {http://adsabs.harvard.edu/abs/2003MNRAS.338..360F} {338, 360}

\bibitem[\protect\citeauthoryear{{Galache}, {Corbet}, {Coe}, {Laycock},
  {Schurch}, {Markwardt}, {Marshall}  \& {Lochner}}{{Galache}
  et~al.}{2008}]{2008ApJS..177..189G}
{Galache} J.~L.,  {Corbet} R.~H.~D.,  {Coe} M.~J.,  {Laycock} S.,  {Schurch}
  M.~P.~E.,  {Markwardt} C.,  {Marshall} F.~E.,   {Lochner} J.,  2008, \mn@doi
  [\apjs] {10.1086/587743}, \href
  {http://adsabs.harvard.edu/abs/2008ApJS..177..189G} {177, 189}

\bibitem[\protect\citeauthoryear{{Gehrels} et~al.,}{{Gehrels}
  et~al.}{2004}]{2004ApJ...611.1005G}
{Gehrels} N.,  et~al., 2004, \mn@doi [\apj] {10.1086/422091}, \href
  {http://adsabs.harvard.edu/abs/2004ApJ...611.1005G} {611, 1005}

\bibitem[\protect\citeauthoryear{{Graczyk}, {Pietrzy{\'n}ski}, {Pilecki},
  {Thompson}, {Gieren}, {Konorski}, {Udalski}  \& {Soszy{\'n}ski}}{{Graczyk}
  et~al.}{2013}]{2013IAUS..289..222G}
{Graczyk} D.,  {Pietrzy{\'n}ski} G.,  {Pilecki} B.,  {Thompson} I.~B.,
  {Gieren} W.,  {Konorski} P.,  {Udalski} A.,   {Soszy{\'n}ski} I.,  2013, in
  {de Grijs} R.,  ed.,  IAU Symposium Vol. 289, Advancing the Physics of Cosmic
  Distances. pp 222--225 (\mn@eprint {arXiv} {1311.1270}),
  \mn@doi{10.1017/S1743921312021436}

\bibitem[\protect\citeauthoryear{{Guillot} et~al.,}{{Guillot}
  et~al.}{2018}]{2018ATel12219....1G}
{Guillot} S.,  et~al., 2018, The Astronomer's Telegram, \href
  {http://adsabs.harvard.edu/abs/2018ATel12219....1G} {12219}

\bibitem[\protect\citeauthoryear{{Harmanec}}{{Harmanec}}{1983}]{1983HvaOB...7...55H}
{Harmanec} P.,  1983, Hvar Observatory Bulletin, \href
  {http://adsabs.harvard.edu/abs/1983HvaOB...7...55H} {7, 55}

\bibitem[\protect\citeauthoryear{{Klus}, {Ho}, {Coe}, {Corbet}  \&
  {Townsend}}{{Klus} et~al.}{2014}]{2014MNRAS.437.3863K}
{Klus} H.,  {Ho} W.~C.~G.,  {Coe} M.~J.,  {Corbet} R.~H.~D.,   {Townsend}
  L.~J.,  2014, \mn@doi [\mnras] {10.1093/mnras/stt2192}, \href
  {http://adsabs.harvard.edu/abs/2014MNRAS.437.3863K} {437, 3863}

\bibitem[\protect\citeauthoryear{{Kobulnicky}, {Nordsieck}, {Burgh}, {Smith},
  {Percival}, {Williams}  \& {O'Donoghue}}{{Kobulnicky}
  et~al.}{2003}]{2003SPIE.4841.1634K}
{Kobulnicky} H.~A.,  {Nordsieck} K.~H.,  {Burgh} E.~B.,  {Smith} M.~P.,
  {Percival} J.~W.,  {Williams} T.~B.,   {O'Donoghue} D.,  2003, in {Iye} M.,
  {Moorwood} A.~F.~M.,  eds,  \procspie Vol. 4841, Instrument Design and
  Performance for Optical/Infrared Ground-based Telescopes. pp 1634--1644,
  \mn@doi{10.1117/12.460315}

\bibitem[\protect\citeauthoryear{{Laycock}, {Corbet}, {Coe}, {Marshall},
  {Markwardt}  \& {Edge}}{{Laycock} et~al.}{2003}]{2003MNRAS.339..435L}
{Laycock} S.,  {Corbet} R.~H.~D.,  {Coe} M.~J.,  {Marshall} F.~E.,  {Markwardt}
  C.,   {Edge} W.,  2003, \mn@doi [\mnras] {10.1046/j.1365-8711.2003.06179.x},
  \href {http://adsabs.harvard.edu/abs/2003MNRAS.339..435L} {339, 435}

\bibitem[\protect\citeauthoryear{{Manick}, {Miszalski}  \& {McBride}}{{Manick}
  et~al.}{2015}]{2015MNRAS.448.1789M}
{Manick} R.,  {Miszalski} B.,   {McBride} V.,  2015, \mn@doi [\mnras]
  {10.1093/mnras/stv074}, \href
  {http://adsabs.harvard.edu/abs/2015MNRAS.448.1789M} {448, 1789}

\bibitem[\protect\citeauthoryear{{Maravelias}, {Antoniou}, {Zezas},
  {Strantzalis}, {Hatzidimitriou}  \& {Haberl}}{{Maravelias}
  et~al.}{2018}]{2018ATel12224....1M}
{Maravelias} G.,  {Antoniou} V.,  {Zezas} A.,  {Strantzalis} A.,
  {Hatzidimitriou} D.,   {Haberl} F.,  2018, The Astronomer's Telegram, \href
  {http://adsabs.harvard.edu/abs/2018ATel12224....1M} {12224}

\bibitem[\protect\citeauthoryear{{Martin}, {Nixon}, {Armitage}, {Lubow}  \&
  {Price}}{{Martin} et~al.}{2014}]{2014ApJ...790L..34M}
{Martin} R.~G.,  {Nixon} C.,  {Armitage} P.~J.,  {Lubow} S.~H.,   {Price}
  D.~J.,  2014, \mn@doi [\apjl] {10.1088/2041-8205/790/2/L34}, \href
  {http://adsabs.harvard.edu/abs/2014ApJ...790L..34M} {790, L34}

\bibitem[\protect\citeauthoryear{{Massey}}{{Massey}}{2002}]{2002ApJS..141...81M}
{Massey} P.,  2002, \mn@doi [\apjs] {10.1086/338286}, \href
  {http://adsabs.harvard.edu/abs/2002ApJS..141...81M} {141, 81}

\bibitem[\protect\citeauthoryear{{McConnachie}}{{McConnachie}}{2012}]{2012AJ....144....4M}
{McConnachie} A.~W.,  2012, \mn@doi [\aj] {10.1088/0004-6256/144/1/4}, \href
  {http://adsabs.harvard.edu/abs/2012AJ....144....4M} {144, 4}

\bibitem[\protect\citeauthoryear{{McCully}, {Volgenau}, {Harbeck}, {Lister},
  {Saunders}, {Turner}, {Siiverd}  \& {Bowman}}{{McCully}
  et~al.}{2018}]{2018SPIE10707E..0KM}
{McCully} C.,  {Volgenau} N.~H.,  {Harbeck} D.-R.,  {Lister} T.~A.,  {Saunders}
  E.~S.,  {Turner} M.~L.,  {Siiverd} R.~J.,   {Bowman} M.,  2018, in Software
  and Cyberinfrastructure for Astronomy V. p. 107070K (\mn@eprint {arXiv}
  {1811.04163}), \mn@doi{10.1117/12.2314340}

\bibitem[\protect\citeauthoryear{{Monageng}, {McBride}, {Coe}, {Steele}  \&
  {Reig}}{{Monageng} et~al.}{2017a}]{2017MNRAS.464..572M}
{Monageng} I.~M.,  {McBride} V.~A.,  {Coe} M.~J.,  {Steele} I.~A.,   {Reig} P.,
   2017a, \mn@doi [\mnras] {10.1093/mnras/stw2354}, \href
  {http://adsabs.harvard.edu/abs/2017MNRAS.464..572M} {464, 572}

\bibitem[\protect\citeauthoryear{{Monageng}, {McBride}, {Townsend}, {Kniazev},
  {Mohamed}  \& {B{\"o}ttcher}}{{Monageng} et~al.}{2017b}]{2017ApJ...847...68M}
{Monageng} I.~M.,  {McBride} V.~A.,  {Townsend} L.~J.,  {Kniazev} A.~Y.,
  {Mohamed} S.,   {B{\"o}ttcher} M.,  2017b, \mn@doi [\apj]
  {10.3847/1538-4357/aa87b7}, \href
  {http://adsabs.harvard.edu/abs/2017ApJ...847...68M} {847, 68}

\bibitem[\protect\citeauthoryear{{Moritani} et~al.,}{{Moritani}
  et~al.}{2013}]{2013PASJ...65...83M}
{Moritani} Y.,  et~al., 2013, \mn@doi [\pasj] {10.1093/pasj/65.4.83}, \href
  {http://adsabs.harvard.edu/abs/2013PASJ...65...83M} {65, 83}

\bibitem[\protect\citeauthoryear{{Rajoelimanana}, {Charles}  \&
  {Udalski}}{{Rajoelimanana} et~al.}{2011}]{2011MNRAS.413.1600R}
{Rajoelimanana} A.~F.,  {Charles} P.~A.,   {Udalski} A.,  2011, \mn@doi
  [\mnras] {10.1111/j.1365-2966.2011.18243.x}, \href
  {http://adsabs.harvard.edu/abs/2011MNRAS.413.1600R} {413, 1600}

\bibitem[\protect\citeauthoryear{{Reig}}{{Reig}}{2011}]{2011Ap&SS.332....1R}
{Reig} P.,  2011, \mn@doi [\apss] {10.1007/s10509-010-0575-8}, \href
  {http://adsabs.harvard.edu/abs/2011Ap%26SS.332....1R} {332, 1}

\bibitem[\protect\citeauthoryear{{Reig} \& {Fabregat}}{{Reig} \&
  {Fabregat}}{2015}]{2015A&A...574A..33R}
{Reig} P.,  {Fabregat} J.,  2015, \mn@doi [\aap] {10.1051/0004-6361/201425008},
  \href {http://adsabs.harvard.edu/abs/2015A%26A...574A..33R} {574, A33}

\bibitem[\protect\citeauthoryear{{Reig}, {Fabregat}  \& {Coe}}{{Reig}
  et~al.}{1997}]{1997A&A...322..193R}
{Reig} P.,  {Fabregat} J.,   {Coe} M.~J.,  1997, \aap, \href
  {http://adsabs.harvard.edu/abs/1997A%26A...322..193R} {322, 193}

\bibitem[\protect\citeauthoryear{{Schwering} \& {Israel}}{{Schwering} \&
  {Israel}}{1991}]{1991A&A...246..231S}
{Schwering} P.~B.~W.,  {Israel} F.~P.,  1991, \aap, \href
  {http://adsabs.harvard.edu/abs/1991A%26A...246..231S} {246, 231}

\bibitem[\protect\citeauthoryear{{Stella}, {White}  \& {Rosner}}{{Stella}
  et~al.}{1986}]{1986ApJ...308..669S}
{Stella} L.,  {White} N.~E.,   {Rosner} R.,  1986, \mn@doi [\apj]
  {10.1086/164538}, \href {http://adsabs.harvard.edu/abs/1986ApJ...308..669S}
  {308, 669}

\bibitem[\protect\citeauthoryear{{Strohmayer} et~al.,}{{Strohmayer}
  et~al.}{2018}]{2018ATel12222....1S}
{Strohmayer} T.~E.,  et~al., 2018, The Astronomer's Telegram, \href
  {http://adsabs.harvard.edu/abs/2018ATel12222....1S} {12222}

\bibitem[\protect\citeauthoryear{{Udalski}, {Kubiak}  \& {Szymanski}}{{Udalski}
  et~al.}{1997}]{1997AcA....47..319U}
{Udalski} A.,  {Kubiak} M.,   {Szymanski} M.,  1997, \actaa, \href
  {http://ukads.nottingham.ac.uk/abs/1997AcA....47..319U} {47, 319}

\bibitem[\protect\citeauthoryear{{Udalski} et~al.,}{{Udalski}
  et~al.}{2008}]{2008AcA....58..329U}
{Udalski} A.,  et~al., 2008, \actaa, \href
  {http://adsabs.harvard.edu/abs/2008AcA....58..329U} {58, 329}

\bibitem[\protect\citeauthoryear{{Udalski}, {Szyma{\'n}ski}  \&
  {Szyma{\'n}ski}}{{Udalski} et~al.}{2015}]{2015AcA....65....1U}
{Udalski} A.,  {Szyma{\'n}ski} M.~K.,   {Szyma{\'n}ski} G.,  2015, \actaa,
  \href {http://ukads.nottingham.ac.uk/abs/2015AcA....65....1U} {65, 1}

\bibitem[\protect\citeauthoryear{{Yang}, {Laycock}, {Christodoulou},
  {Fingerman}, {Coe}  \& {Drake}}{{Yang} et~al.}{2017}]{2017ApJ...839..119Y}
{Yang} J.,  {Laycock} S.~G.~T.,  {Christodoulou} D.~M.,  {Fingerman} S.,  {Coe}
  M.~J.,   {Drake} J.~J.,  2017, \mn@doi [\apj] {10.3847/1538-4357/aa6898},
  \href {http://adsabs.harvard.edu/abs/2017ApJ...839..119Y} {839, 119}

\makeatother
\end{thebibliography}



\end{document}